\documentclass[10pt,onecolumn,preprintnumbers]{revtex4}
\pdfoutput=1

\usepackage[english]{babel}
\usepackage[latin1]{inputenc}
\usepackage{graphicx}
\usepackage{amsmath}
\usepackage{color}
\usepackage{ulem}

\begin{document}

\title{Ellipsoidal Universe and Cosmic Shear}
%
%

\author{Luigi Tedesco}
\email{luigi.tedesco@ba.infn.it}
\affiliation{Istituto Nazionale di Fisica Nucleare, Sezione di Bari, Bari, Italy}
\affiliation{Dipartimento di Fisica, Universit\`a di Bari, Via G. Amendola 173, 70126, Bari, Italy}

\date{\today}

\begin{abstract}
We consider a Bianchi I geometry of the Universe. We obtain a cosmic shear expression related with the eccentricity of the Universe. In particular we study the connection among cosmic shear, eccentricity and CMB. The equation are self-contained with only two parameters. 

\end{abstract}


\maketitle 

{\bf 1. Introduction}
\\
\\
\\

The standard model of the universe shows great compatibility with the observational data. The homogeneity and isotropy of the universe seem to be two acquired aspects in the so-called standard model of the universe. Clearly, much depends on the scales we are examining. When we consider local scales, the universe is neither homogeneous nor isotropic, and, generally, galaxies, galaxy clusters, and large-scale structures such as cosmic filaments and voids create anisotropies. On cosmological scales, these anisotropies must be averaged, and the universe becomes homogeneous and isotropic over about 100~megaparsecs. But, despite all the successes, this model shows several dark spots, the nature of dark matter and a dark energy, the Hubble tension, the coincidence  and fine-tuning problem, and so on.

There is often a great deal of confusion about this in the sense that one thing is the anisotropy of the universe at the ``local'' level and another thing is the anisotropy of the universe in the early universe. We discuss here the second anisotropy.

In this paper, we consider a Bianchi I model of the universe that describes a homogenous and anisotropic universe. There is much evidence regarding the anisotropy of the universe, and the possibility of the anisotropy of our universe is not an academic discussion. The origins of the study in this sector are in 1933 and {1968} \cite{lemaitre,jacobs1}, and the works are very interesting and pioneering. To be fair, it was Einstein who directed Lemaitre to study these anisotropic universes. The first classification is due to Luigi Bianchi in his famous classification (Lie algebra is the first classification) \cite{bianchi}.  

Anisotropy is an important aspect in relation to the evolution of the universe; for example, the  anomalies in the CMB indicate that we have a plane-mirroring symmetry in large scales  \cite{gurza1,gurza2,buchert,schwarz,russell}, or the large-angle anomalies in the CMB scale
\cite{bennet}, or  the low-quadrupole moment in the CMB \cite{barrow1,berera,campanelli1,campanelli2,copij,Rodrigue} in which it is possible to indicate an ellipsoidal expansion of the universe because the low-quadrupole moment of the CMB is suppressed at large scales. In fact, the CMB data put in evidence a suppression of the power when we consider large angular scales, the so-called quadrupole problem.  Another aspect of this model is the large-scale asymmetry  \cite{eriksen,hansen} or a very strange cold spot \cite{vielva} or the {}alignment of quadrupole and octupole moments 
 \cite{land,raltston,deoliveira,copi}. A very interesting recent work regards the study of more precise observational bounds on global anisotropy \cite{loeb}. Among other things, this work studies how the difference between the different expansion rates disappears as the universe evolves, eventually showing a universe in which the differences in anisotropy disappear, demonstrating an isotropic universe in evolution. In any case, many interesting works have been conducted regarding the anisotropic universe, for example,  \cite{aluri,tedesco,tedesco2,akarsu,campanelli,quercellini,sarmah,campa,deliduman,Mukherjee,deandrade}, and, conversely, in this context regarding an ellipsoidal universe. Indeed, several aspects have been studied, including the parallax effect \cite{pippo2}. The importance of an ellipsoidal universe consists of considering this model as a valid alternative to the FLRW models if we mean small deviations from the perfect isotropy.  

{A very interesting study consists of connecting the anisotropy in the context of the Finslerian structure of space--time because, when we consider these spaces, we are able to investigate the phenomena of the local anisotropy connected with a gravitational field and the cosmology   \cite{stavrinos1,stavrinos2,stavrinos3,stavrinos4,stavrinos5,stavrinos6,gonner,stavrinos7}}.

The mechanisms that enable the universe to become anisotropic are different, such as a uniform cosmological magnetic field \cite{campanelli1,campanelli2} or topological defects (such as cosmic strings or domain walls) \cite{berera}. On the other hand, it is not possible to neglect motions whose origin is due to  inhomogeneity and local anisotropy  \cite{sarkar}. A very interesting aspect regards  the new data of the Pantheon supernova from the late universe; these data  show new hints of anisotropy \cite{zhao,Amirhashch,colin}.
In this paper, we analyze cosmic shear, a powerful concept to explore the distribution of matter in the universe, the understanding of the cosmology, and the evolution of the large-scale structures. We will connect the shear with the eccentricity of the ellipsoidal universe. In this work, we also analyze the deep connections between cosmic shear, eccentricity, and CMB, as evident in the last section.  

{The importance of considering the CMB is linked to a double factor; the first is the fact that the temperature of the cosmic microwave background does not provide a total isotropy, but, with the precision of the order of $10^{-5}$, the second factor is the problem of large-angle anomalies and the quadrupole problem, i.e., the discrepancy between the theoretical prediction for l = 2 and the experimental one, which are in deep disagreement. Therefore, understanding the connection between the anisotropy of the universe and the CMB is not a purely academic exercise but can be a good starting point for understanding and decoding those mechanisms that are still unknown and that hide profound connections.}

The paper is organized as follows. In Section~\ref{sec:2}, we discuss the fundamental properties of an ellipsoidal universe. In Section~\ref{sec:3}, we analyze the evolution equation for anisotropic energy density. In Section~\ref{sec:4}, we study the cosmic shear with an ellipsoidal universe.  The quadrupole of the CMB, shear, and eccentricity are investigated in Section~\ref{sec:5}. We summarize and discuss our results in the final section.
\vskip 2truecm
{\bf 2. Model of the ellipsoidal Universe}
\\
\\
\\
The line element, spatially homogeneous and anisotropic Bianchi type-I, is described by the most general plane-symmetric line element, the so-called Taub line element \cite{taub}     
\begin{equation}
\label{metric}
ds^2 = dt^2 - a^2(t) (dx^2+dy^2) - b^2(t) dz^2
\end{equation}
where $a$ and $b$ are two time-dependent different scale factors along two different axes of space, $x \, y$ and $z$, respectively. Equation {(\ref{metric})} describes a space in which the expansion is~ellipsoidal.

Bianchi I model may be considered an intriguing alternative to FLRW model of the universe if we consider small deviations from the isotropy.

A very interesting relation between these anisotropic models and quadrupole, octupole, and limit on the shear, vorticity, and Weyl tensor has been considered in \cite{maart}.
For completeness, we write the   non-vanishing Christoffel symbols
\begin{equation}
\Gamma^0_{11}=\Gamma^0_{22} = a \, \dot{a}, \,\,\,\,\,\,\,\,\,\,\,\,\,\, \Gamma^0_{33}=b \, \dot{b},
\,\,\,\,\,\,\,\,\,\,\,\, \Gamma^1_{01}=\Gamma^2_{02} = \dot{a}/a, \,\,\,\,\,\,\,\,\,\,\,\,\,\,\,\,\,\,\,\, \Gamma^3_{03} = \dot{b}/b
\end{equation}
where  $\dot{ } \equiv d/dt$ is the derivative with respect to the cosmic time. The non-vanishing Ricci tensor components are
\begin{equation}
R^0_{0} = -2 \left( \frac {\ddot{a}} {a} + \frac {\ddot{b}} {b} \right), \,\,\,\,\,\,\,\,\,\,\, R^1_1 = R^2_2 = - \left[ \frac {\ddot{a}} {a} + \left( \frac {\dot{a}} {a} \right)^2 + \frac {\dot {a} \dot {b}} {a b} \right], \,\,\,\,\,\,\,\,\,\,\,
R^3_3 = - \left( \frac {\ddot {b}} {b} + 2 \, \frac {\dot {a} \, \dot{b}} {a \, b} \right).
\end{equation}
The energy--momentum tensor, considering planar symmetry, is
\begin{equation}
T^{\mu}_{\,\, \nu} = \text{{diag}} \,  (\rho, -p_{\parallel}, -p_{\parallel} -p_{\perp})
\end{equation}
with $\rho$ energy density, $p_{\parallel}$ longitudinal pressure, and $p_{\perp}$ transversal pressure. This tensor, in an anisotropic universe, is provided by the contributions of the sum of two components, one isotropic and one anisotropic. The anisotropic component is
\begin{equation}
(T_A)^{\mu}_{\,\, \nu} = \text{{diag}} \, (\rho^A, -p^A_{\parallel}, -p^A_{\parallel}, -p^A_{\perp})
\end{equation}
and an isotropic contribution is given by
\begin{equation}
(T_I)^{\mu}_{\,\, \nu} = \text{ {diag}} (\rho^I, -p^I, -p^I, -p^I)
\end{equation}
where it can be obtained by means of the three components: radiation (r), matter(m) and cosmological constant ($\Lambda$):
\begin{equation}
\rho^I= \rho_r + \rho_m+ \rho_{\Lambda}
\end{equation}
\begin{equation}
p^I= p_r + p_m + p_{\Lambda}
\end{equation}
where 
$p_r=\rho_r/3, \,\,\, p_m=0, \,\,\, p_{\Lambda} = \rho_{\Lambda}$. Different authors studied  the exact solutions of  Einstein's equations in relation to different plane-symmetric and isotropic components~\cite{berera,berera2,barrow}. 
\\
Let A be the mean expansion parameter
\begin{equation}
A\equiv (a^2 b)^{1/3}, \,\,\,\,\,\,\,\,\, H_a= \frac {\dot{a}} {a}, \,\,\,\,\,\,\,\,\, H_b= \frac {\dot{b}} {b}
\end{equation}
with $H_a$ and $H_b$ the directional Hubble parameters along the directions  (x,y) and z, respectively.  
The average expansion rate or ``mean Hubble parameter'' is provided by
\begin{equation}
H= \frac {\dot{A}} {A} = \frac{2 H_a + H_b} {3}
\end{equation}
and the spatial volume $V$ is 
\begin{equation}
V = \sqrt{ -g} = a^2b \, .
\end{equation}
Another important element is the so-called ``scalar expansion'' or ``expansion rate'', indicated by $\theta$ and  defined as
\begin{equation}
\theta = 3H = 3 \frac {\dot {{A}}} {A}  
\end{equation}
while the scalar shear is \cite{tedesco,akarsu,maurya}
\begin{equation}
{\sigma}^2 = \frac {1} {2} \sigma_{\alpha \beta} \sigma^{\alpha \beta} =\frac {1} {2}  \left( 2 H_a^2 + H_b - \frac {\theta^2} {3} \right)
\end{equation}
where $\sigma_{\alpha \beta} = \frac {1} {2} (u_{\mu;\nu} + u_{\nu;\mu}) {h^{\mu}}_{\alpha} {h^{\nu}}_{\beta} - \frac {1} {3} u^{\mu}_{; \mu} h_{\alpha \beta}$ is called shear tensor and $h_{\mu \nu} = g_{\mu \nu} + u_{\mu} u_{\nu}$ is the projection tensor, while $u_{\mu}$ is the four-velocity in the comoving coordinates.
The components of the cosmic shear vector $\vec{\Sigma}$ are 
\begin{equation}
\Sigma_{x,y,z} \equiv \frac {H_{x,y,z} -H} {H}\, .
\end{equation}
Therefore, 
\begin{equation}
\label{sigma}
\Sigma_a= \frac {H_a-H} {H}, \,\,\,\,\,\,\,\,\,\, \Sigma_b= \frac {H_b-H} {H}, \,\,\,\,\,\,\, \text{and}   \,\,\,\,\,\,\, \Sigma_a= \frac {H_a - H_b} {2 H_a + H_b}. 
\end{equation}
From
 a physical point of view, the shear or the scalar expansion indicates how the anisotropic universe is.

Now, we want to connect the eccentricity, `$e$, of the universe with $\Sigma$. Let us first define the eccentricity as
\begin{equation}
e= \sqrt{1 - \frac {b^2} {a^2}}  \phantom{sddhwei} \,\,\,\, \text{if} \,\,\,\,\,\, a>b   \phantom{sddiiwei} {or}  \phantom{sddhwwei} e= \sqrt{1 - \frac {a^2} {b^2}}  \phantom{sdhdwei} \,\,\,\,\, \text{if}  \,\,\,\,\,\,\,b>a.
\end{equation}
{We} normalize the eccentricity in order to obtain  $a(t_0)=b(t_0)=1$.

Let us consider the first case, ($a>b$); we have 
\begin{equation}
b^2 = a^2 (1- e^2).
\end{equation}
{If we} derive both sides by dt and divide for $b^2$, we obtain
\begin{equation}
2 \, \frac {\dot b} {b} = 2 \, a \, \dot{a} \frac {(1 - e^2)} {b^2} + \frac {a^2} {b^2} (- 2 \, e \, \dot{e})
\end{equation}
from which we have the important connection between $H$ and $e$
\begin{equation}
H_b=H_a - \frac  {e \, \dot{e}} {1 - e^2}
\end{equation}
(for $e = \sqrt{1 - \frac {b^2} {a^2}} $). Of course, the other case,  $e = \sqrt{1 - \frac {a^2} {b^2}} $, provides $H_a=H_b - \frac  {e \, \dot{e}} {1 - e^2}$. For completeness, we write important relations
\begin{equation}
\label{HHa} 
H_a= H \pm \frac {e \, \dot{e}} {3 (1-e^2)}
\end{equation}
and 
\begin{equation}
\label{HHb} 
H_b= H \mp \frac {2} {3} \frac  {e \, \dot{e}} { (1-e^2)}
\end{equation}
where the signs ``+'' and ``$-$'' in Equation (\ref{HHa}) refer to $e=\sqrt{1-b^2/a^2}$ and $e=\sqrt{1-a^2/b^2}$, respectively.

Combining  Equations (15) and (19), we can derive an interesting relation linking $\Sigma$ and $e$:
\begin{equation}
\label{kkk}
\Sigma_a = \frac {e \, \dot {e}}  {3 H (1- e^2)} \,\,\,\,\,\,\,\,\,\, \text{for} \,\,\,\,\,\,\,\,\,\,\,\,\,\, e= \sqrt{1 - \frac {b^2} {a^2}} 
\end{equation}
and of course $\Sigma_b = - \frac {e \, \dot {e}}  {3 H (1- e^2)}$ for $e= \sqrt{1 - \frac {a^2} {b^2}}$.
Now, we derive the evolution equation for the eccentricity $e$.
Let us start with Einstein's equations, in a Bianchi I universe
\begin{eqnarray}
\left\{
\begin{array}{rl}
&
\left(\frac {\dot{a}} {a} \right)^2 + 2 \frac {\dot{a} \,  \dot{b}}  {a \, b} = 8 \pi G \,(\rho^I + \rho^A) \\ \\
&
\frac {\ddot{a}} {a} + \frac {\ddot{b}} {b} + \frac {\dot{a} \,  \dot{b}}  {a \, b} = - 8 \pi G \, (p^I + p^A_{\parallel}) \\ \\
&
2 \frac {\ddot{a}} {a} + \left(\frac {\dot{a}} {a} \right)^2 = -8 \pi G \, (p^I + p^A_{\perp})
\end{array}
\right.
\end{eqnarray}
Taking into account that $\frac {\ddot{a}} {a} = \dot{H_a} + {H_a}^2 $, it is possible to write these equations in terms of~$H_{a,b}$:
\begin{eqnarray}
\label{eqdiH}
\left\{
\begin{array}{rl}
&
H_a^2+ 2H_a H_b = 8 \pi G \rho \\ \\
& 
{\dot{H}_a} + H_a^2 + {\dot{H}}_b + H^2_b + H_a H_b = - 8 \, \pi \, G \, p_{\parallel} \\ \\
&
2 {\dot{H}}_a + 3 H_a =  - 8 \, \pi \,  G \, p_{\perp},
\end{array}
\right.
\end{eqnarray}
with $\rho=\rho^I + \rho^A$, \, $p_{\parallel} = p^I +p_{\parallel}^A$ and \, $p_{\perp} = p^I + p_{\perp}^A$. Subtracting the second from the third equation in Equation (24), we obtain 
\begin{equation}
\label{qwe}
\dot{H_a} + 2 H_a^2 - \dot{H_b} - H_b^2 - H_a H_b = 8 \pi G (p_{\parallel} - p_{\perp}).
\end{equation}
{If we set $E= \frac {e \dot{e}} {3 (1 -  e^2)}$, we have 
$
H_a= H+ E  \,\text{and}\,   H_b = H - 2E,$ which resolve Equation~(\ref{qwe}) as}
\begin{equation}
3 \dot{E} + 9 H E = 8 \pi G (p_{\parallel} - p_{\perp}),
\end{equation}
from which we finally have the most general temporal evolution for the eccentricity:
\begin{equation}
\label{ecc}
\frac {d} {dt} \left( \frac {e \, \dot{e}} {1 - e^2}  \right) + 3 H \left( \frac {e \, \dot{e}} {1 - e^2} \right) = \pm \,  8 \pi G \, (p_{\parallel} - p_{\perp})
\end{equation}
where "+" refers to the eccentricity $e = \sqrt{1-b^2/a^2}$ and the sign "-" refers to $e = \sqrt{1-a^2/b^2}$. Eq.(\ref{ecc}) is the exact equation, which can be conveniently simplified by neglecting second-order terms. E' importante sottolineare che e' la piu  genrale espressione di evoluzione nel tempo della eccentricita'. It is important to underline that it is the most general expression of the evolution of the eccentricity over time. Clearly the resolution of the differential equation must take into account the boundary condition that $e(t_0)=0$ that is to say the Universe today is isotropic.
\vskip 1.5 truecm
{\bf 3. Evolution equation for anisotropic energy density}
\\
\\
We use the conservation of the anisotropic component of the energy momentum tensor provided by Equation (4)
\begin{equation}
\label{cons}
(T^A)^\mu_{\,\, \nu; \mu} = \frac {\partial T_{\nu}^\mu} {\partial x^{\mu}} + \Gamma^{\mu}_{\mu \lambda} T^{\lambda}_{\nu} - \Gamma^{\lambda}_{\mu \nu} T^{\mu}_{\lambda} = 0.
\end{equation}
Let us consider the only component $\nu=0$ of the conservation of the anisotropic energy momentum tensor, $(T^A)^\mu_{\,\, 0; \mu}$, and the non-vanishing terms are
\begin{equation}
\mu=0 \,\,\,\,\rightarrow \,\,\,\, \frac {\partial {T_0}^0} {\partial x^{0}} + \Gamma^{0}_{0 \lambda} T^{\lambda}_{0} - \Gamma^{\lambda}_{0 0} T^{0}_{\lambda} = \dot{\rho}^A
\end{equation}
\begin{eqnarray}
\mu=1 \,\,\,\,\rightarrow \,\,\,\, \frac {\partial {T_0}^1} {\partial x^{1}} + \Gamma^{1}_{1 \lambda} T^{\lambda}_{0} - \Gamma^{\lambda}_{1 0} T^{1}_{\lambda} \,\,\,\, \rightarrow \,\,\,\,
\label{eqdiH}
\left\{
\begin{array}{rl}
&
\text{for} \,\,\,\, \lambda = 0 \,\,\,\,  \text{we have} \,\,\,\, \frac {\dot{a}} {a} \rho^A \\ \\
& 
\text{for} \,\,\,\, \lambda =1 \,\,\,\, \text{we have} \,\,\,\, \frac  {\dot{a}} {a} p^A_{\parallel} \\ \\
\end{array}
\right.
\end{eqnarray}
\begin{eqnarray}
\mu=2 \,\,\,\,\rightarrow \,\,\,\, \frac {\partial {T_0}^2} {\partial x^{2}} + \Gamma^{2}_{2 \lambda} T^{\lambda}_{0} - \Gamma^{\lambda}_{2 0} T^{2}_{\lambda} \,\,\,\, \rightarrow \,\,\,\,
\left\{
\begin{array}{rl}
&
\text{for} \,\,\,\, \lambda = 0 \,\,\,\,  \text{we have} \,\,\,\, \frac {\dot{a}} {a} \rho^A \\ \\
& 
\text{for} \,\,\,\, \lambda =2 \,\,\,\, \text{we have} \,\,\,\, \frac  {\dot{a}} {a} p^A_{\parallel} \\ \\
\end{array}
\right.
\end{eqnarray}
\begin{eqnarray}
\mu=3 \,\,\,\,\rightarrow \,\,\,\, \frac {\partial {T_0}^3} {\partial x^{3}} + \Gamma^{3}_{3 \lambda} T^{\lambda}_{3} - \Gamma^{\lambda}_{3 0} T^{3}_{\lambda} \,\,\,\, \rightarrow \,\,\,\,
\left\{
\begin{array}{rl}
&
\text{for} \,\,\,\, \lambda = 0 \,\,\,\,  \text{we have} \,\,\,\, \frac {\dot{b}} {b} \rho^A \\ \\
& 
\text{for} \,\,\,\, \lambda =3 \,\,\,\, \text{we have} \,\,\,\, \frac  {\dot{b}} {b} p^A_{\perp} .\\ \\
\end{array}
\right.
\end{eqnarray}
Finally, putting all together, Equation (\ref{cons}) provides the 00 component of the conservation of the energy momentum anisotropic tensor:
\begin{equation}
\label{consrho}
{\dot{\rho}}^A + 2 \, \frac{\dot{a}} {a} (\rho^A + p^A_{\parallel}) + \frac {\dot {b}} {b} \, (\rho^A + p^A_{\perp}) =0.
\end{equation}
From a cosmological point of view, it is important to find how density evolves. To this end, we put Equations (\ref{HHa}) and~(\ref{HHb}) in Equation (\ref{consrho})
\begin{equation}
\label{rhof}
\dot{\rho}^A + 2 \left(H \pm \frac {e \, \dot{e}} {3 ( 1- e^2)} \right) (\rho^A + p^A_{\parallel} ) + \left(H \mp \frac {2} {3} \frac {e \, \dot{e}} {1- e^2} \right) (\rho^A + p^A_{\perp})
\end{equation}
and after a few steps we have 
\begin{equation}
\label{pippo}
{\dot{\rho}}^A + H \, (3 \rho^A + 2 p^A_{\parallel} + p^A_{\perp}) \pm \frac {2} {3} \frac {e \, \dot{e}} {1 - e^2} (p^A_{\parallel} - p^A_{\perp}) =0.
\end{equation}
Now, if we neglect the last term because $p^A_{\parallel} \simeq  p^A_{\perp}$ and
if we assume a classical barotropic relation between pressure and density, that is to say
$p^A_{\parallel} = w^A_{\parallel} \rho_A$ and $p^A_{\perp}=w^A_{\perp} \rho^A$,\linebreak Equation~(\ref{pippo}) becomes
\begin{equation}
\dot{\rho}^A + H \rho^A (2 + 2 w^A_{\parallel} + w_{\perp})=0
\end{equation}
which is a first-order differential equation with the initial condition $\rho^A(t=0) = \rho^{A}_{(0)}$, and we have the following solution
\begin{equation}
\label{rhosol}
\rho^A= \rho^{A}_{(0)} \frac {1} {A^{3+2w^A_{\parallel}+w^A_{\perp}}}.
\end{equation}
The eq.(\ref{rhosol}) is interesting because
we are expressing the anisotropic energy density (which is an elusive quantity) of the universe by the anisotropic energy density now and the barotropic relations involving $w_{\parallel}$ and $w_{\perp}$ (see for example \cite{{pippo1},{pippo2}}).
\vskip 1truecm
{\bf 4. Cosmic shear and ellipsoidal Universe}
\\
\\
In this section we want to connect the cosmic shear to the  parameter that most characterizes an ellipsoidal universe, namely the eccentricity $e$. To do this we start from the most general eccentricity evolution equation, eq.(\ref{ecc}), and introduce the shear, eq.(\ref{kkk}). Let us put $L(t) \equiv H(t) \, \Sigma(t)$ we have
\begin{equation}
\frac {dL} {d t} + 3 Hy= \frac {8 \pi G} {3} (p^A_{\parallel} - p^A_{\perp}).
\end{equation} 
The general solution is 
\begin{equation}
L(t) = \frac {1} {A^3} \left[ \int d t' \, \frac {8 \pi G} {3} ( w^A_{\parallel} - w^A_{\perp}) \, \rho^A A^3 +cost \right] \equiv \frac {1} {A^3} [M(A) + const]
\end{equation}
where $const$ is the integration constant. Now, $t_0$ is the actual time,  $L(t_0) = M[A({t_0}) / A^3(t_0)] + cost /A^3(t_0)$ but $A(t_0)=1$, this implies $L(t) = 1/ A^3 [ M(A(t)) + L(t_0) - M(t_0)]$, in other terms we can write (remembering eq.(\ref{kkk}))%
\begin{equation}
L(t) = \frac {1} {A^3} \left[ H_0 \Sigma_a + \Omega_A^{(0)} H_0 ( w_{\parallel}^A - w_{\perp}^A )\int_1^A  \frac {d \xi} {\xi^{1+2 w_{\parallel}^A + w^A_{\perp}} \, H/H_0} \right] = \frac {e \, \dot{e}} {3 (1 - e^2)}
\end{equation}
Let us observe that $1-e^2 \simeq 1$ and $2 \, e \, \dot{e} = de^2 / dt$, therefore
\begin{equation}
\label{ddd}
\frac {d e^2} {d t} = \frac {6 H_0} {A^3} \left[ \Sigma_a + \Omega_A^{(0)} H_0 ( w_{\parallel}^A - w_{\perp}^A )\int_1^A  \frac {d \xi} {\xi^{1+2 w_{\parallel}^A + w^A_{\perp}} \, H/H_0} \right].
\end{equation}
Now Integrate eq.(\ref{ddd}) between $t_0$ time today and $t_{dec}$ time at decoupling, and let us observe that $\int_{t_0}^{t_{dec}} dt = \int_1^{A_{t_{dec}}} \frac {dA} {A H}$
\begin{equation}
\int_{t_0}^{t_{dec}} \frac {d e^2} {dt} = e^2_{dec} = 6 H_0 \left[ \Sigma_a \int_{1}^{A_{dec}} \frac {dA} {A^4 H}+ \int_1^{A_{dec}} \Omega_A^{(0)} H_0 (w_{\parallel}^A - w_{\perp}^A ) \int_1^{A_{{dec}}}  \frac {d \xi} {{\xi^{1+2 w_{\parallel}^A + w^A_{\perp}}} \, H^2 /H_0} \frac {dA} {A^4} \right],
\end{equation}
and dividing both members by $e^2$ we obtain an equation of this type
\begin{equation}
1 = c_1 \, \Sigma_a + c_2 \, \Omega^{(0)}_A.
\end{equation}
The two unknowns are $\Sigma_a$ and $\Omega^{(0)}_A$, we have managed to simplify the evolution equations in the simplest way by connecting sigma and omega. Clearly we have one equation and two unknowns, it is necessary to obtain another equation.
\vskip 1.5truecm
{\bf 5. Quadrupole of CMB and cosmic shear}
\\
\\
In this section we want to obtain the second relation in order to fix one of the  two important parameter of the previous section.  In order to find this equation, we rewrite the 
metric eq.(\ref{metric}), taking into account that the eccentricity in small 
\begin{equation}
\label{metr1}
ds^2 = dt^2 - A^2 (t) \, (\delta_{ij} + h_{ij}) \, d x^i dy^j.
\end{equation}
The null geodesic equation in a perturbed Friedmann-Lemaitre-Robertson-Walker metric gives the temperature anisotropy (Sachs-Wolfe effect)
\begin{equation}
\label{delta}
\frac {\Delta T} {<T>} =  - \frac {1} {2} \int_{t_0}^{t_{dec}} dt \, \frac {\partial h_{ij}} {\partial t} n^i n^j
\end{equation}
where $n^i$  are the directional cosines. Let's find the value of $h_{ij}$ starting with eq.(\ref{metric}) that can be written as 
\begin{equation}
\label{metr2}
ds^2 = dt^2 - A^2(t) \left[ \delta_{ij} + \frac {a^2} {A^2} \delta_{ij} + \frac {a^2 - b^2}  {A^2}  \delta_{i3} \delta_{j3} - \delta_{ij} \right] dx^i dx^j  .
\end{equation}
From the comparison with the two equations, eq.(\ref{metr1}) with  eq.(\ref{metr2}), we obtain 
\begin{equation}
\label{acca}
h^{ij} = \left( \frac {a^2} {A^2} -1 \right) \delta^{ij} + \frac {a^2 - b^2} {A^2} \delta^{i3} \, \delta^{j3}.
\end{equation}
Since in eq.(\ref{delta})  we have the time derivative of $h_{ij}$ we derive with respect to time eq.(\ref{acca})
\begin{equation}
\frac {\partial h^{ij}} {\partial t} = 2 \frac {a^2} {A^2} (H_a - H) \, \delta^{ij} + 2 \frac {a^2} {A^2} \left[ H_a - \frac {b^2} {a^2} H_b - H \left(1 - \frac {b^2} {a^2} \right) \right] \delta^{i3} \delta^{j3} .
\end{equation}
Now remembering that $H_a - H = H \,  \Sigma_a$ we have
\begin{equation}
\frac {\partial h^{ij}} {\partial t} = 2 \, \frac {a^2} {A^2} \, H \, \Sigma_a \, \delta^{ij} + 2 \frac {a^2} {A^2} \left[ H \, \Sigma_a - \frac {b^2} {a^2} \Sigma_b \right] \delta^{i3} \delta^{j3}
\end{equation}
but $\Sigma_b= -2 \Sigma_a$ therefore
 \begin{equation}
 \label{www}
\frac {\partial h^{ij}} {\partial t}= 2 \, \frac {a^2} {A^2} \, H \, \Sigma_a \left[ \delta^{ij} + \left(1 + 2 \frac {b^2} {a^2} \right)  \delta^{i3} \delta^{j3} \right].
\end{equation} 
Let's go back to the starting equation eq.(\ref{delta}), substituting in it the eq.(\ref{www}), we finally have
\begin{equation}
\frac {\Delta T} {<T>} = -  \int_{t_0}^{t_{dec}} dt \, \frac {a^2} {A^2} \, H \, \Sigma_a \, \left[ 1 + \left( 1 + 2 \frac {b^2} {a^2} \right) n_3^2 \right]
\end{equation}
where $ \delta^{ij} n^i n^j =1$ and $\delta^{i3} \, \delta^{j3} \, n^i \, n^j = n_3^2$ 
We have that the two equations eq.(43) and eq.(51) identify our two unknowns. An indicative value for $\Omega_A$ can be deduced from the equation  (2.20) of \cite{campanelli2} where 
\begin{equation}
e_{dec} \simeq 10^{-2} \sqrt{ \Omega_A /10^{-7}}.
\end{equation}
Now imposing that now the Universe is isotropic $e(t_0)=0$, from the evolution equation of the eccentricity eq.(27) we have that at decoupling ($z=1088$) $e_{dec}=10^{-2}$. In other terms $\Omega_A \simeq 10^{-7}$ that is of course a first order of magnitude of this unknown quantity, that is not zero. Another observation is connected to the eq. (43) because if we take into account this value of $\Omega$, we have that $ \Sigma_a  \simeq k_1 - k_2 \Omega_A$, in other terms are fundamental the determination of the constants that can also be orders of magnitude. We have only two parameters to be played in the next.
\\
A final qualitative analysis regards eq.(51) that  provides a useful equation for determining our unknowns. The observed quadrupole anisotropy at CMB gives $(\Delta T_2)_{obs} \simeq 250 \mu K^2$, therefore considering the equation that relates the temperature variation of CMB with the eccentricity (see eq.(3.13) of \cite{campanelli}) $\Delta T / <T> = - 1/2 \, e^2_{dec} n^2_3$ it is possibile to obtain $n^2_3$ in fact from  the inverse formula  $n^2_3= -2 \Delta T / <T> 1/ e_{dec}^2$. We set $\Delta T= \sqrt{250 \times 10^{-6}} K, <T> \simeq 2.7 K$ and $ e_{dec} \simeq 10^{-2}$  we obtain $n_3^2 \sim 120$ that is a value that can be utilized in eq.(51). 
\vskip 1.5truecm
{\bf 6. Conclusions}
\\
\\
In this paper, we have investigated the connections among the eccentricity of an ellipsoidal universe, the lower quadrupole of the large-scale CMB anisotropy data, and the cosmic shear. 

A non-negligible aspect is that, in the literature, either anisotropic or non-homogeneous models of the universe are almost always considered  separately. In reality, taking both models (at the same time) into consideration is not an ``error'' because the cosmological principle model is only valid on a very large scale. There are works that consider models of the universe that simultaneously take into account anisotropy and non-homogeneity, such as \cite{fanizza}, in which anisotropic and inhomogeneous models of the universe are studied. In this paper, we study a universe in which inhomogeneities and anisotropies coexist and thus obtain Einstein's equation. This approach will enable further investigations regarding universes that are simultaneously both non-isotropic and non-homogeneous, providing variations with respect to considering only one of the two aspects with related properties. It is probable that considering only one aspect (or only isotropy or only non-homogeneity) makes us lose useful information and probably hides properties that would emerge with the considered model.

Different energy densities, synthesized in $\rho^A$, can contribute to the Hubble parameter in a different weighted way. Therefore, the anisotropic nature of the early universe provides a very interesting energy content in relation to H. This analysis has been completed in recent works \cite{binici,deliduman}  that analyze the discrepancy between the existence of distant galaxies that are very large, primordial, and, therefore, early, compared to a $\Lambda$CDM model that would provide a star formation too slowly.

Equations (43) and (51) are important  because no constraint exists on $\Sigma_a$ in the literature and an order of magnitude of $\Sigma_a \sim 10^{-5}$ \cite{campanelli}.

A very important consideration regards ``isotropization'' because, generally, when we consider an anisotropic universe, the Einstein gravity and the cosmological constant tend to  cancel the anisotropy; in other terms, the universe becomes isotropic asymptotically. 

Practically,  a way to realize the isotropization consists of considering the limit for directional expansion factor $a= b$ for $t \rightarrow \infty $; on the other hand, it is possible to obtain the isotropization with \cite{jacobs,bronnikov,saha1,saha2} the mean expansion parameter $A \rightarrow 0$ and the scalar shear  $\sigma^2 \rightarrow 0$.

 This work is a completion of \cite{tedesco}. There are substantial differences between the two works because, in \cite{tedesco}, we try to frame the physics using the CS to relate everything to the accelerating and decelerating phases of the universe. This aspect has not been addressed here, just as the correlation between the variation in acceleration of the universe, the so-called jerk parameter, eccentricity, and cosmic shear have not been addressed. In this work, we have instead tried to link the cosmic shear to the eccentricity passing through the cosmic microwave background. Interestingly, from a temporal point of view, these three ``indicators'' are coexisting: we are at about z = 1080. What is new is to correlate a link between these parameters and the variation in the temperature of the CMB related to the average $\frac {\Delta T}{T}$. We have therefore arrived at Equation  (51), which, with Equation (43), represent the final synthesis of all the reasoning and enable it to be a good starting point for new developments.

 As regards the comparison with \cite{russell}, there are profound differences. While in this work the pressure is practically the same in the three different expansion axes, in this work, a pressure is introduced parallel and perpendicular to the expansion axes, which is not only different from a calculation point of view but conceptually profound. 
 The work of~\cite{russell} focuses the analysis on the isotropization of the universe to make it FLRW, all due to a sort of contraction of the scale factors, a process that would allow us to move from a Bianchi I universe to an FLRW universe. For the moment, in this work, this aspect has been put aside because it will be the subject of a work \cite{inpreparation} not linked to any contraction but as an evolution of an external factor (a magnetic field, for example). In~\cite{russell}, the authors find, on the anisotropic model, some constraints that are provided by the upper limits of anisotropy from the Planck data. In addition, in that work, isotropization criteria are introduced. In this work, however, the possible experimental constraints are only those related to the position of the quadrupole (l = 2), leaving the cosmic shear free for the moment,  although Equation (43) can determine important constraints on the cosmic shear since $\Omega_A^{(0)}$ will have to be derived from the Friedmann equation.
On the other hand, in this work, instead, the eccentricity of the universe is linked to the cosmic microwave background through the parameter of cosmic shear, which considers things from another point of view.  A determining role is played by eccentricity and its evolution in time, as can be seen from Equation (27). From this point of view, the approach with respect to the work of \cite{russell} is completely different since it sheds light on the close connection between eccentricity and cosmic shear and their temporal evolution.

The limits of this model are strictly connected to the transition between the anisotropic universe and the current isotropic universe. In other words, it is necessary to connect the origin of the anisotropy to the causes that generated this ellipsoidal universe. These causes may later disappear by ``washing out'' the anisotropy itself. An example is the primordial magnetic field that can be a good generator of anisotropy \cite{parker,barrow}. 

Following the expansion of the universe, the primordial magnetic fields become less and less intense, so they act less and less. Today, traces of these cosmic magnetic fields remain in the universe, whose origin could be connected to the anisotropy of the universe itself. This is a limit of the model itself because it takes little to eliminate the seed magnetic fields, but, at the same time, it is a potential area of research for future developments. An aspect often overlooked in these models, and which could provide great contributions to the understanding of the universe, is this: if the universe expanded like an ellipsoid, it means that one axis expanded more rapidly than the other. Why was there this disparity in expansion? Think of the shape of an egg, in which one axis is larger than the other, generating the characteristic shape because space was stretched more in one direction than the other. Answering these apparently simple questions means entering the realm of the mysteries of the universe. Another important evolution in these anisotropic models can be connected with the current $H_0$, in which we have a deep discrepancy between the early time and late time determinations of the Hubble constant. On the other hand, the ellipsoidal model of the universe may solve the so-called $S_8$ tension \cite{pcea}. Another aspect, as already mentioned, concerns how to isotropize this ellipsoidal universe. It is not at all trivial. There are new studies that approach this problem in a very promising way. The delicate problem concerns the need to find the cosmological limits that enable testing them with experimental data \cite{inpreparation}.

It is beyond the scope of this work, but, for completeness, next, we will study the problem of isotropizing the model, which does not simply mean reducing the eccentricity to zero. As already demonstrated in \cite{tedesco}, the isotropization must take into account several factors, such as $\Sigma$ , as well as the jerk parameter, which we have not dealt with in this work, a parameter that must be set to 1 in the $\Lambda$CDM model. 
On the other hand it is essential to connect these reasonings to the CMB polarization data, to have a better connection with the experimental data \cite{cea}.
\\
In \cite{inpreparation}, the study combines cosmic shear with eccentricity and two important cosmological parameters: the acceleration parameter (which is essentially connected with the second derivative of the expansion parameter)  and the jerk (which is essentially connected with the third derivative of the expansion parameter). Preliminary studies indicate that there is a strong correlation between these parameters that are not simply linked by standard relations but provide interesting indicators on their evolution, strictly connected to eccentricity. The study must consider that, if there is a universe with two expansion parameters, like the one considered in this work, it is necessary to consider two acceleration parameters and two jerk parameters that, in the end, when the universe returns to being isotropic, must provide a single contribution. Last but not least, it is interesting to study the luminosity distance in relation to an ellipsoidal universe, where the value of the eccentricity is not negligible.
In other terms, considering an ellipsoidal universe is not a painless process for modern cosmology, which cannot ignore it.
\\
\\
\\
\\
{\bf Acknowledgments}
he work is  supported in part by INFN under the program TAsP: ``Theoretical Astroparticle Physics'' and also this work was partially supported by the research grant number 2022E2J4RK ``PANTHEON: Perspectives in Astroparticle and
Neutrino THEory with Old and New messengers'' under the program PRIN 2022 funded by the Italian Ministero dell?Universit\'a e della Ricerca (MUR) and by the European Union-Next Generation EU. I want to acknowledge the hospitality of CERN, Theoretical Department, where this work has been carried out.
\\
\\
 \end{document}